\documentclass[aps,preprint,showpacs,preprintnumbers,amsmath,amssymb,floatfix]{revtex4}
\usepackage[german,english]{babel}
\usepackage{bm,amsmath,amsfonts,dsfont}
\usepackage{amssymb}
\usepackage{graphicx}
\usepackage{subfigure}
\usepackage{xcolor}

\def\beq{\begin{equation}}
\def\eeq{\end{equation}}
\def\beqn{\begin{eqnarray}}
\def\eeqn{\end{eqnarray}}

\begin{document}

\title{Dynamics and symmetries of a repulsively bound atom pair in an infinite optical lattice} 

\author{Andreas~Deuchert$^{1}$\footnote{Corresponding author. E-mail: Andreas.Deuchert@pci.uni-heidelberg.de},
Kaspar~Sakmann$^{1}$,
Alexej~I.~Streltsov$^{1}$,  
Ofir~E.~Alon$^{2}$,
and Lorenz~S.~Cederbaum$^{1}$}

\affiliation{$^1$ Theoretische Chemie, Physikalisch-Chemisches Institut, Universit\"at Heidelberg,\\
Im Neuenheimer Feld 229, D-69120 Heidelberg, Germany}

\affiliation{$^2$ Department of Physics, University of Haifa at Oranim, Tivon 36006, Israel}

\begin{abstract}
We investigate the dynamics of two bosons trapped in an infinite one-dimensional optical lattice potential within the framework of the Bose-Hubbard model and derive an exact expression for the wavefunction at finite time. As initial condition we chose localized atoms that are separated by a distance of $d$ lattice sites and carry a center of mass quasi-momentum. An initially localized pair ($d=0$) is found to be more stable as quantified by the pair probability (probability to find two atoms at the same lattice site) when the interaction and/or the center of mass quasi-momentum is increased. For initially separated atoms ($d \neq 0$) there exists an optimal interaction strength for pair formation. Simple expressions for the wavefunction, the pair probability and the optimal interaction strength for pair formation are computed in the limit of infinite time. Whereas the time-dependent wavefunction differs for values of the interaction strength that differ only by the sign, important observables like the density and the pair probability do not. With a symmetry analysis this behavior is shown to extend to the $N$-particle level and to fermionic systems. Our results provide a complementary understanding of the recently observed [Winkler \textit{et al.}, Nature (London) \textbf{441}, 853 (2006)] dynamical stability of atom pairs in a repulsively interacting lattice gas.   
\end{abstract}

\maketitle

\section{Introduction}
The physics of particles trapped in periodic potentials has been a topic of extensive research since the early days of quantum theory \cite{Wannier1937, Kohn1959, Ashcroft}. The progress in experimental techniques during the last two decades which has led to the realization of atomic Bose-Einstein condensates (BECs) in optical lattice potentials and the possibility to tune the inter-atomic interaction via Feshbach resonances has renewed this interest \cite{RevModPhys.80.885, Lewenstein2007, yukalov2009}. Because of the high control over the system's parameters and the absence of strong dissipation channels, it is possible to simulate periodic systems isolated from other effects like, for example, phononic degrees of freedom which play an important role in solid state physics. Due to this isolation not only ground state properties but also the excited states and the dynamics of such systems play a crucial role in the understanding of present experiments. 

Of special interest for this work is the experiment recently done by Winkler \textit{et al.} \cite{Repnature2006} who show that two repulsively interacting atoms initially prepared at one site of an optical lattice potential separate \textit{less} rapidly than their non-interacting counterpart. After this experiment several theoretical works on the topic followed, most of them within the framework of the Bose-Hubbard model. Different two-body problems (with several trapping potentials and interactions) are investigated in \cite{valiente_states, valientequantumdynamics2008, Javanainen2010, wang2008, PhysRevA.78.023617, valientescattering2009, PhysRevA.81.042102, PhysRevA.81.045601}. Other authors study the effect using a larger ensemble of particles \cite{petrosyan-2007-76, PhysRevA.79.063634}. But pairing induced by a repulsive interaction is not restricted to bosonic atoms, it is a relevant topic for fermions as well \cite{PhysRevA.80.015601, PhysRevA.78.063617, valiente2010}. A work that does not use the framework of a Hubbard or Bose-Hubbard model is an extension of an older work on fermionic pairing in the context of high temperature superconductivity by Mahajan and Thyagaraja \cite{Mahajan2006, Mahajan1997}, who point out that the effect of pairing by repulsion has three ingredients, namely quantum mechanics, a periodic potential and a short range interaction. 

In this work we are interested in pairing by repulsion (or attraction) from the viewpoint of dynamics. Usually the computation of quantum dynamics is a difficult task and a hot topic of actual research \cite{Pethick2006, stringari2003, RevModPhys.83.863, PhysRevLett.99.030402,PhysRevA.77.033613}. Because of the complexity of the problem approximations are often a must and most computations are based upon numerical methods. Therefore, exactly solvable models are of special interest. In this work we present an exact solution of the two-body dynamics of two initially localized atoms carrying a center of mass quasi-momentum within the framework of the Bose-Hubbard model. The solution is then applied to the physical problem of pairing of two bosonic atoms trapped in an optical lattice potential. Common observables are found to depend only on the magnitude of the interaction strength but not on its sign. With a symmetry analysis this behavior is shown to extend to the corresponding (bosonic or fermionic) $N$-particle system. The paper is structured as follows. In Sect.~\ref{sec:dynamics} we derive the time-dependent wavefunction and discuss results of the dynamics at finite time. A simple way to solve the dynamics in the limit of infinite time is presented in Sect.~\ref{sec:asymptotic_dynamics}. The following section (Sect.~\ref{sec:symmetry_analysis}) is concerned with the question how the dynamics of the pair and some generalization for the $N$-particle system depend on the sign of the interaction strength. Finally, in Sect.~\ref{sec:summary} we summarize the findings and give a short outlook.
\section{Dynamics at finite time}
\label{sec:dynamics}
\subsection{The time-dependent wavefunction}
\label{subsec:wavefunction}
The Bose-Hubbard model mostly is used in its second quantized version \cite{PhysRevB.40.546}. The Hamiltonian describing particles in an infinite one-dimensional periodic potential reads
\begin{equation}
\hat{H} = -J \sum_{\alpha \in \mathbb{Z}} \left( \hat{b}_{\alpha}^{\dagger} \hat{b}_{\alpha+1} + \hat{b}_{\alpha+1}^{\dagger} \hat{b}_{\alpha} \right) + \frac{U}{2} \sum_{\alpha \in \mathbb{Z}} \hat{n}_\alpha \left( \hat{n}_{\alpha} - 1 \right).
\label{eq:dynamics1}
\end{equation}
As usual $\hat{b}_{\alpha}^{\dagger}$/$\hat{b}_\alpha$ is the creation/annihilation operator and $\hat{n}_{\alpha}$ the number operator for a particle at site $\alpha$, which is described by a Wannier function centered around the lattice site. $J$ and $U$ denote the tunneling rate between neighboring sites and the on-site interaction, respectively. When treating two-particle systems within the Bose-Hubbard model its representation in first quantization is more convenient \cite{Repnature2006, daley2006, one-dimensional-hubbard}. The two-particle Schr\"odinger equation is then given by the expression
\begin{equation}
-J \left[ \Delta_x + \Delta_y \right] \Psi(x,y) + U \delta_{x,y} \Psi(x,y) = E \Psi(x,y).
\label{eq:dynamics2}
\end{equation}  
The coordinates $x$ and $y$ are elements of the lattice $\Gamma = \mathbb{Z}$ (lattice constant $a=1$) and for the Laplace operator on the lattice we define $\Delta_x \Psi(x) = \Psi(x+1) + \Psi(x-1)$. To exploit the translational symmetry we introduce relative and center of mass coordinates $r = x-y$, $R = \frac{x+y}{2}$. The center of mass coordinate is an element of the lattice $\Gamma' = \left( \mathbb{Z}/2 \right)$, while the relative coordinate is an element of the original lattice $\Gamma$. For the wavefunction the usual ansatz $\Psi(x,y) = e^{i K R} \Psi_K(r)$ can be made. The time-independent Schr\"odinger equation thus transforms to
\begin{equation}
\left[ -J_K \Delta_r + U \delta_{r,0} \right] \Psi_K(r) = E_K \Psi_K(r) \quad \text{for} \quad K \in \left[ -\pi, \pi \right],
\label{eq:dynamics3}
\end{equation}
where $J_K = 2 J \cos(K/2)$ is an effective hopping parameter. Please see \footnote{The possible values of $K$ are derived as follows. Let $T$ be the translation operator that shifts both coordinates, $T \Psi(x,y) = \Psi(x+1,y+1)$. Since $\left[ T, H \right] = 0$ it can be diagonalized simultaneously with the Hamiltonian $H$. The spectrum of $T$ is given by the unit circle in the complex plane, hence if we use the ansatz $\Psi(R,r) = e^{i K R} \Psi_K(r)$ we have to choose $K \in \left[ -\pi, \pi \right]$.} for an argument why $K \in \left[ -\pi, \pi \right]$ although the center of mass lattice $\Gamma'$ has a lattice constant of $a_{CM} = \frac{1}{2}$. The solutions of the above eigenvalue equation consist of one bound state and a continuum of scattering states for each value of $K$ \cite{valiente_states, Repnature2006}. In contrast to the corresponding continuum model with delta function interaction \cite{FliesbachQM}, the bound state exists also for repulsive interaction. For the bound state wavefunction and energy there are two distinct expressions, one for attractive interaction and one for repulsive interaction
\begin{align}
\Psi_K^{BS}(R,r) &= 
\begin{cases}
\frac{e^{i K R}}{\sqrt{2 \pi}} \frac{ \sqrt{\left| U_K \right|}}{\sqrt[4]{U_K^2+1}} \left( \sqrt{U_K^2+1} - \left| U_K \right| \right)^{\left| r \right|}, $ for $  U < 0, \\
\frac{e^{i K R}}{\sqrt{2 \pi}} \frac{ \sqrt{U_K}}{\sqrt[4]{U_K^2+1}} \left( U_K  - \sqrt{U_K^2+1} \right)^{\left| r \right|}, $ for $  U > 0,
\end{cases} \\
E_{BS}(K) &= 
\begin{cases}
-\sqrt{U^2 + 4J_K^2}, $ for $  U < 0, \\
 \sqrt{U^2 + 4J_K^2}, $ for $  U > 0.	
\end{cases} 
\label{eq:dynamics4}
\end{align}	
$U_K = U/(2J_K)$ denotes an effective interaction parameter. We note that the bound state has the same density $\rho^{BS}_K(R,r) = \left| \Psi^{BS}_K (R,r) \right|^2$ for repulsive and for attractive interaction. Its energy lies below the scattering continuum for $U < 0$ (lowest state in energy) and above it for $U > 0$ (highest state in energy). The scattering states appear in the literature in an unnormalized form. We provide here the normalized functions
\begin{align}
\Psi_{K,k} (R,r) &= \frac{ \frac{1}{ \sqrt{2 \pi} } e^{i K R} }{ \sqrt{ \pi \left( 1 + \frac{U_K^2}{\sin(k)^2} \right) } } \left\{ \cos(k r) + \frac{U_K}{\sin(k)} \sin(k \left| r \right|) \right\}, \\
E(K,k) &= -2J_K \cos(k), 
\label{eq:dynamics5}
\end{align}
$k \in \left[-\pi, \pi \right]$. Note the symmetry of the spectrum for $K \rightarrow -K$ and $k \rightarrow -k$. 

Given an initial state at $t=0$ the dynamics of the system can be computed with the time evolution operator $\hat{U}(t) = e^{-i \hat{H} t}$ ($\hbar = 1$). Making an expansion in the eigenbasis of the Hamiltonian this calculation can be done explicitly. As an initial condition we make the choice $\Psi_{0,0}(R,r) = \frac{e^{i Q R}}{\sqrt{2 \pi}} \delta_{r,0}$ to describe two particles sitting at the same lattice site and having a center of mass quasi-momentum $Q$, or $\Psi_{0,d}(R,r) = \frac{e^{i Q R}}{\sqrt{2 \pi}} \frac{1}{\sqrt{2}} \left( \delta_{r,d} + \delta_{r,-d} \right)$ for two particles that are initially separated by a distance of $d$ lattice sites. Expanding the initial state into the above basis and propagating it in time we find the time-dependent wavefunction
\begin{align}
\Psi_{U \leq 0}( R, r, t ) &= \frac{e^{i Q R}}{\sqrt{2 \pi}} \left[ c_d I(r,t) + c_d \frac{ e^{i \sqrt{U^2 + 4 J_{Q}^2} t} \left| U_Q \right|}{\sqrt{1+U_Q^2}} \left( \sqrt{1+U_Q^2} - \left| U_Q \right| \right)^{\left| r \right| + \left| d \right|} \right], 
\label{eq:dynamics6a} \\
\Psi_{U \geq 0}( R, r, t ) &= \frac{e^{i Q R}}{\sqrt{2 \pi}} \left[ c_d I(r,t)  + c_d \frac{ e^{-i \sqrt{U^2 + 4 J_{Q}^2} t} U_Q}{\sqrt{U_Q^2+1}} \left( U_Q  - \sqrt{U_Q^2+1} \right)^{\left| r \right| + \left| d \right|} \right].
\label{eq:dynamics6b}
\end{align} 
$c_d$ is a normalization constant which equals one for $d=0$ and $\sqrt{2}$ for $d \neq 0$. The function $I(r,t)$ is given by
\begin{align}
I(r,t) &= \int_{0}^{\pi} \frac{dk}{\pi} e^{i 2 J_Q t \cos(k)} \frac{f_d(k) f_r(k)}{1 + \frac{U_Q^2}{\sin(k)^2}},
\label{eq:dynamics7a} \\
f_n(k) &= \cos(k n) + \frac{U_Q}{\sin(k)} \sin(k \left| n \right|),
\label{eq:dynamics7b} 
\end{align}
where $n \in \mathbb{Z}$. It is a closed form integral expression that can easily be evaluated numerically with standard computer algebra programs. In what follows we use the notation 
\begin{equation}
\Psi(R,r,t)=\frac{e^{i Q R}}{ \sqrt{2 \pi}} \Phi_Q(r,t)
\label{eq:dynamics8}
\end{equation} 
to have an explicit expression for the time-dependent wavefunction in the relative coordinate at hand. We note that $\Phi_Q(r,t)$ stands for the expression in brackets in Eq.~\eqref{eq:dynamics6a} [Eq.~\eqref{eq:dynamics6b}] for $U \leq 0$ ($U \geq 0$). Since the wavefunction in the center of mass coordinate is a spatially oscillating exponential function, the two-particle density depends only on the relative coordinate.
\subsection{Dynamics of the pair probability and the density in the relative coordinate}
\label{subsec:quantities}
We are interested in the question how a pair forms or dissociates in time. To quantify the process we use the pair probability which measures the probability to find two particles at the same lattice site \cite{PhysRevA.81.045601}. The pair probability is of special interest because it is one of the observables used in the experiment on repulsively bound atom pairs \cite{Repnature2006}. For a two-particle system it can be defined as the expectation value of the pair operator $\hat{P}_{pair} = \sum_{\alpha \in \mathbb{Z}} \left| n_{\alpha} = 2 \right\rangle \left\langle n_{\alpha} = 2 \right|$ which in first quantization reads $\hat{P}_{pair} = \delta_{r,0}$. Therefore, the pair probability equals the density in the relative coordinate at point $r=0$,
\begin{equation}
P_{pair}(t) = \left| \Phi_Q(r = 0,t) \right|^2.
\label{eq:dynpairprob1}
\end{equation}
Our second observable is the variance of the pair probability which we use to measure its fluctuations. We note that the pair operator is a projector and fulfills the relation $\hat{P}_{pair}^2 = \hat{P}_{pair}$. Hence, the variance of the pair probability can be written as a function of the pair probability itself: 
\begin{equation}
\sigma^2_{pair}(t) = \left\langle \left( \hat{P}_{pair} - \left\langle \hat{P}_{pair} \right\rangle \right)^2 \right\rangle = P_{pair}(t) \left[ 1 - P_{pair}(t) \right].
\label{eq:dynpairprob2}
\end{equation}
As a third observable we monitor the density in the relative coordinate in order to get an impression how the initial wave packet changes its shape. It is given by
\begin{equation}
\rho(r,t) = \left| \Phi_Q(r,t) \right|^2.
\label{eq:dynpairprob3}
\end{equation}
All calculations have been carried out for $J=1$ and $Q=0$. Other values of the hopping parameter and the center of mass quasi-momentum scale the time $t$ linearly with $J_K = 2 J \cos(K/2)$ and the on-site interaction $U$ inversely with $J_K$ \footnote{Assume the Hamiltonian from Eq.~\eqref{eq:dynamics3}. When we define $\hat{H}' = \hat{H}/J_K = -\Delta + \frac{U}{J_K} \delta_{r,0}$ and $t' = J_K t$ the time evolution operator can be written as $\hat{U}(t) = e^{-i \hat{H}' t'}$. Now it is clear that the time $t$ scales linearly with $J_K= 2 J \cos(K/2)$ and the interaction strength $U$ inversely.}. 

The three computed quantities, $P_{pair}(t)$, $\sigma^2_{pair}(t)$ and $\rho(r,t)$ depend only on the magnitude of the interaction strength and not on its sign. Therefore, it is sufficient to treat the case $U \geq 0$. We note that this behavior is a nice explanation for the existence of repulsively bound atom pairs. Using the initial conditions from Sect.~\ref{sec:dynamics}, the repulsively interacting particles act as if they would attract each other and vice versa. Hence, there is a strong relation between the phenomena of binding by repulsion and binding by attraction. In Sect.~\ref{sec:symmetry_analysis} we will derive a similar result for the $N$-particle system which gives a direct link to the experiment on repulsively bound atom pairs.

\begin{figure}[htb]
\centering
{
\subfigure[]{
\includegraphics[scale=0.225]{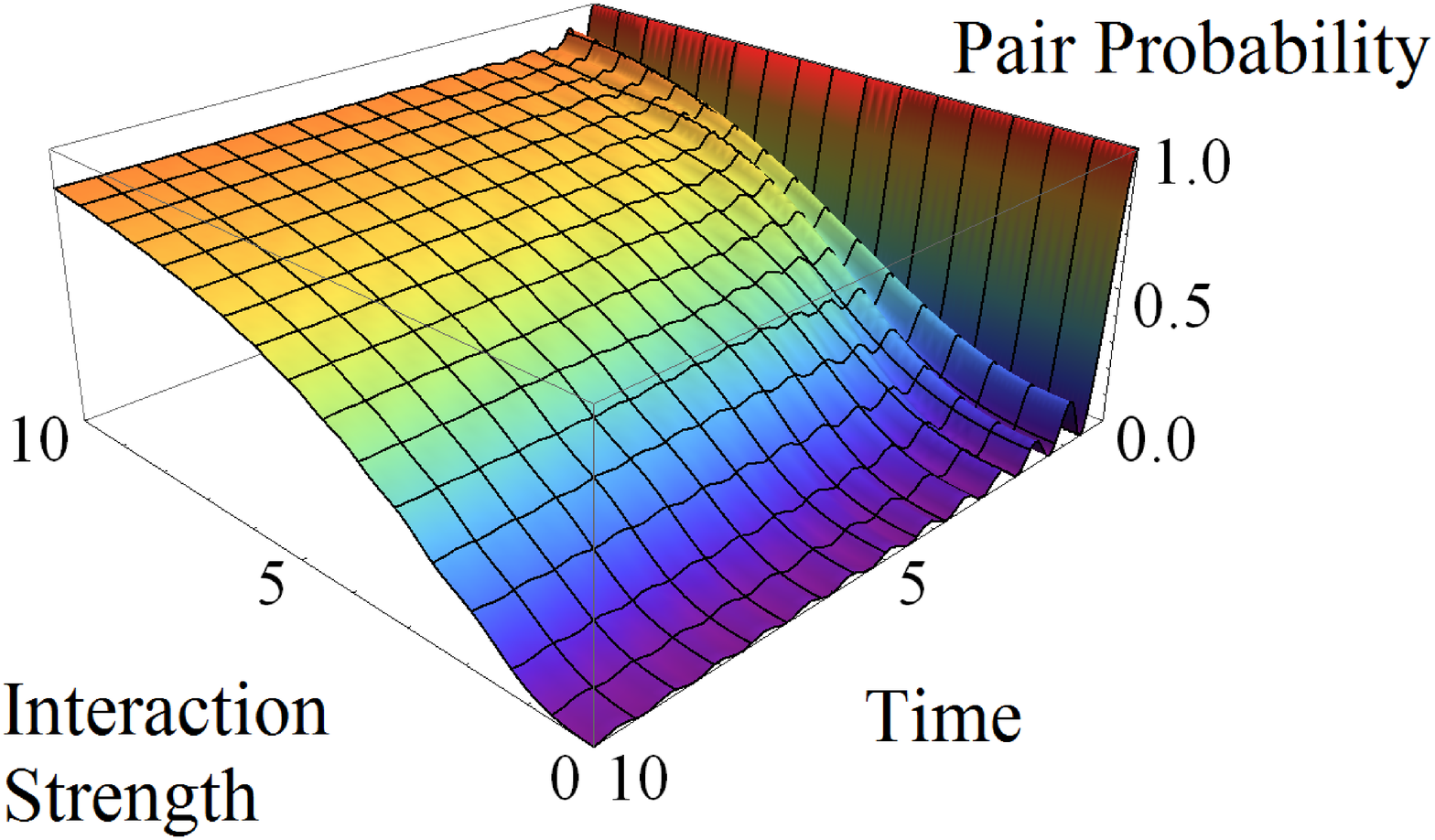}
\label{fig:pairprob3d1a}
}
\subfigure[]{
\includegraphics[scale=0.225]{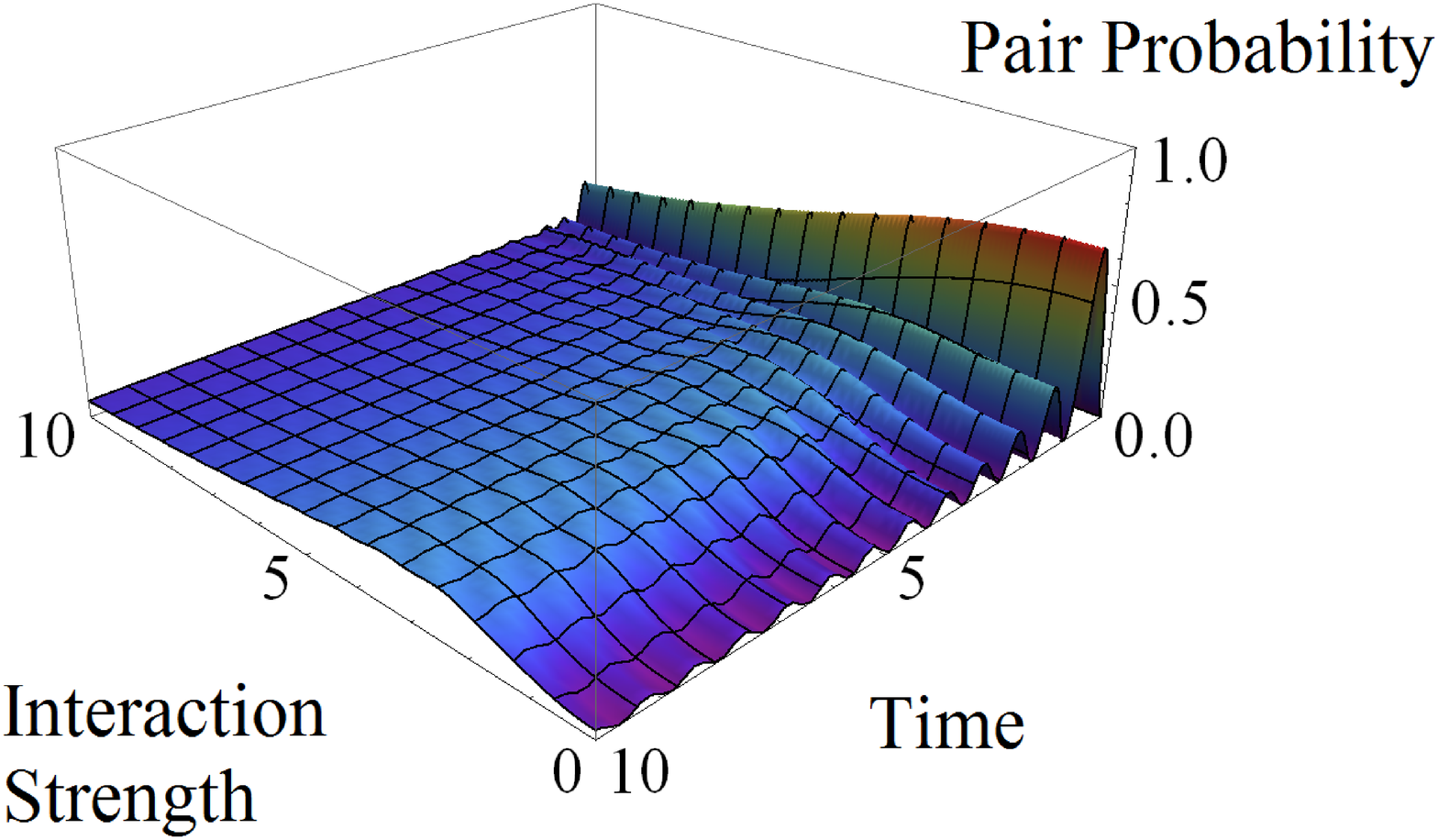}
\label{fig:pairprob3d1b}
}
\caption{(Color online) Pair probability $(P_{pair})$ as a function of the interaction strength $(U)$ and time $(t)$ for $J=1$, $K=0$, $d=0$ (a) and $d=1$ (b). With increasing time $P_{pair}$ is converging rapidly towards an asymptotic value. For $d=0$ the pair probability increases with the interaction strength while for $d=1$ there exists an optimal interaction strength for pair formation (except for very short times). See Sect.~\ref{subsec:quantities} for more details. All quantities are dimensionless.}
\label{fig:pairprob3d1}
}
\end{figure}

When we start the dynamics with two atoms localized at the same lattice site [see Fig.~\ref{fig:pairprob3d1a}] the pair probability starts at $P_{pair}(t=0) = 1$. For $U=0$ it oscillates for small times and then rapidly goes to zero as $1/t$ for large $t$. For $U \neq 0$ this behavior changes. Again, we have some oscillations for small times but for large $t$ the pair probability evolves towards a finite asymptotic value. This asymptotic value is reached more quickly in case of a larger interaction strength. Additionally, the asymptotic value of the pair probability is found to increase with the interaction strength. For an arbitrary fixed time the pair probability is an increasing function of the interaction strength. Hence, we conclude that if we start the dynamics with a pair (two particles at the same lattice site), a large interaction strength stabilizes the pair. Since the effective interaction strength scales inversely with $\cos(Q/2)$ \footnotemark[\value{footnote}] also a large center of mass quasi-momentum stabilizes an initially prepared atom pair (for $U \neq 0$).

If we start the dynamics with two atoms sitting at adjacent lattice sites [$d = 1$, $P_{pair}(t=0) = 0$] the qualitative shape of the pair probability changes although some phenomenology stays the same [see Fig.~\ref{fig:pairprob3d1b}]. In the non-interacting case ($U=0$) we again have some oscillations for small times, and again $P_{pair}(t)$ goes to zero for large times. Additionally, like for $d=0$ the pair probability evolves towards a finite asymptotic value as long as $U \neq 0$. An important difference is that it has a maximum as a function of $U$ for all finite $t$ (except for very small times). The existence of this maximum can be understood as an interplay between two effects in the process of pair creation. As long as the two particles are separated, a small interaction strength is needed for pair formation. But when the particles have come together and are sitting at the same lattice site, a large interaction strength is needed to stabilize the pair. Hence, there exists an optimal interaction strength for pair formation which we have computed analytically in the limit of infinite time, see Sect.~\ref{sec:asymptotic_dynamics}. 

We note that the argumentation used to explain the existence of an optimal interaction strength for pair formation holds for repulsive as well as for attractive interaction. In the latter case a strong attractive interaction keeps the atoms apart from each other which is counterintuitive at first sight. The effect can be understood in the following way. Two atoms coming together at one lattice site lower their interaction energy and due to energy conservation they gain additional kinetic energy. When the width of the first Bloch band ($4 J_Q$) is smaller compared to $U$ this process is strongly suppressed.

\begin{figure}[htb]
\centering
{
\subfigure[]{
\includegraphics[scale=0.235]{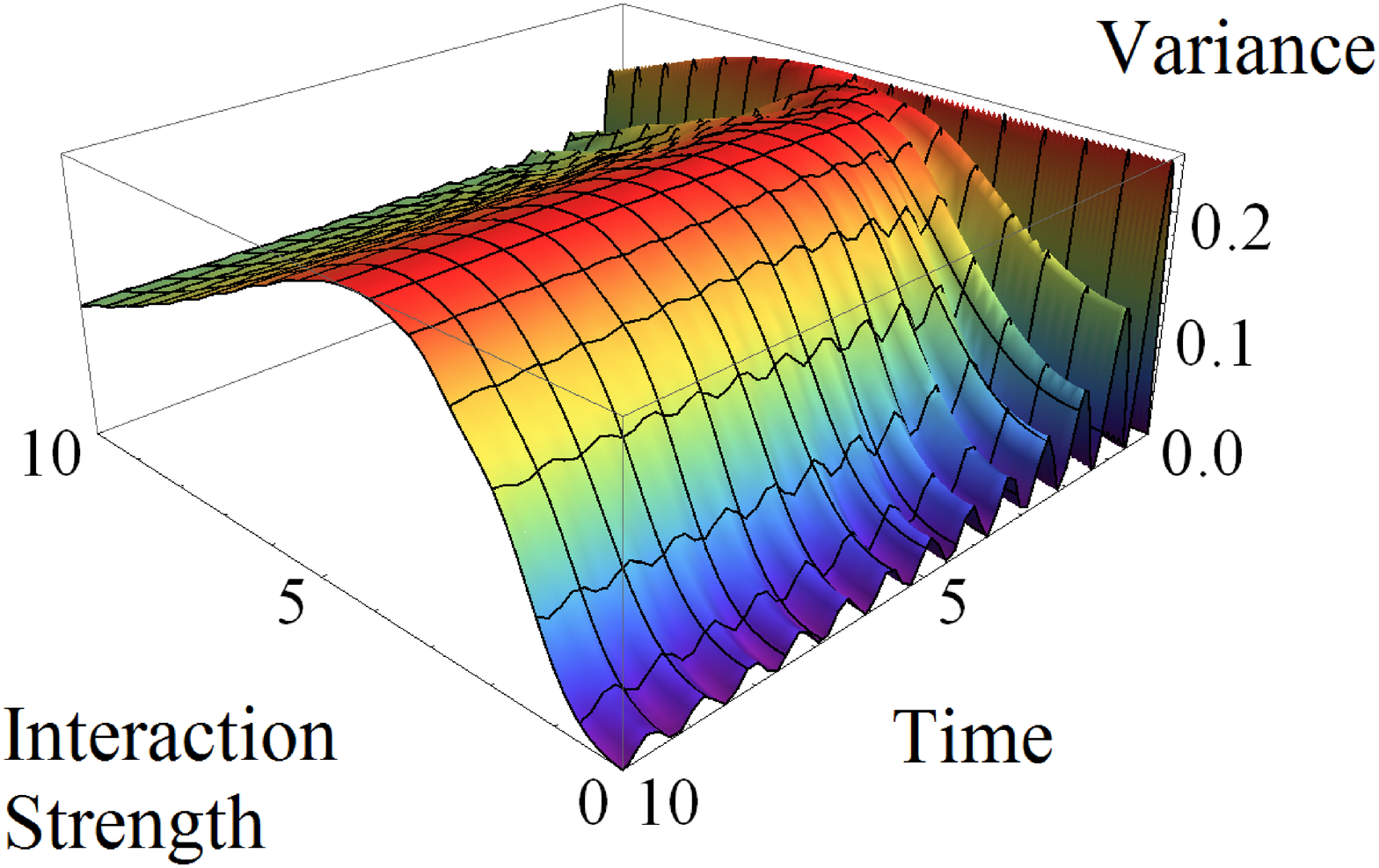}
\label{fig:varaince3d1a}
}
\subfigure[]{
\includegraphics[scale=0.235]{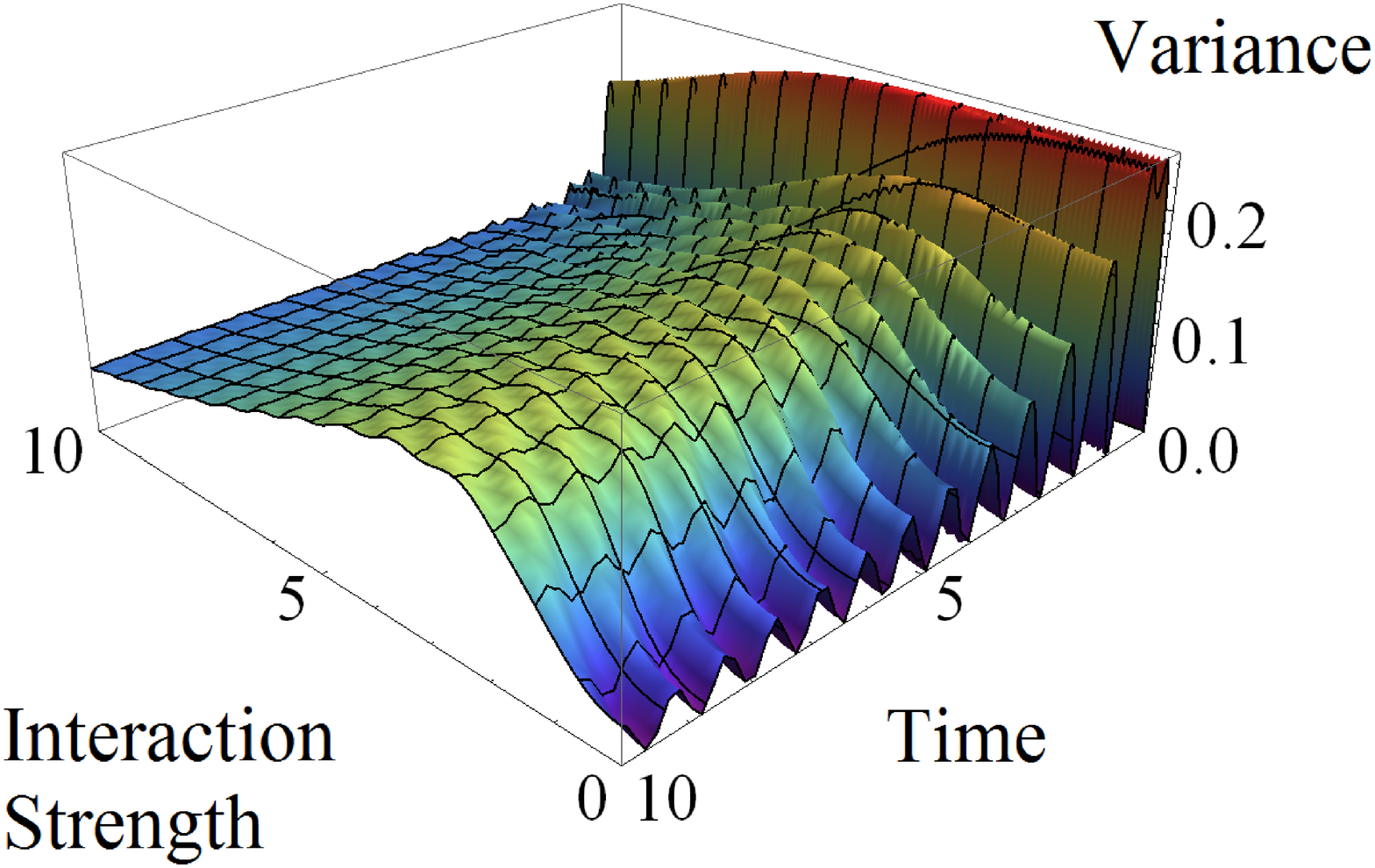}
\label{fig:variance3d1b}
}
\caption{(Color online) Variance $(\sigma_{pair})$ of the pair probability as a function of the interaction strength $(U)$ and time $(t)$ for $J=1$, $K=0$, $d=0$ (a) and $d=1$ (b). Both curves have very similar characteristics, namely a local maximum for intermediate interaction strengths and the convergence towards a constant value for large times. For $d=1$ the maximum of the pair probability and its variance coincide, see Sect.~\ref{subsec:quantities} for more details. Therefore, in this regime a large value of the pair probability goes hand in hand with large fluctuations. All quantities are dimensionless.}
\label{fig:variance3d1}
}
\end{figure}

The variance of the pair probability as a function of the pair probability itself $\sigma_{pair}^2(t) = P_{pair}(t) \left[ 1 - P_{pair}(t) \right]$ is symmetric around $P_{pair}(t) = \frac{1}{2}$ where it takes its maximum; and for $P_{pair}(t) = 0, 1$ it equals zero. We conclude that the variance is small whenever the pair probability is near to zero or one. Large fluctuations occur only in case of intermediate pair probabilities. For $d=0$ and $d=1$ the variance of the pair probability plotted as a function of $U$ and $t$ has very similar characteristics, see Fig.~\ref{fig:variance3d1}. For $U = 0$ both curves start at $\sigma_{pair}^2(t) = 0$, then they quickly increase, oscillate and go to zero for large times. Like in the case of the pair probability, this behavior changes when a finite interaction strength is turned on. The two surfaces now converge to an asymptotic value larger than zero. For all fixed and not too small times $t$ ($t > 2$ is sufficient), both functions possess a local maximum. In case of $d=1$ this maximum coincides with the maximum of the pair probability since the latter mentioned is smaller than $\frac{1}{2}$ for all $U$ and $t$, and hence in this regime a large pair probability always goes hand in hand with large fluctuations of this quantity. 

\begin{figure}[htb]
\centering
{
\includegraphics[scale=0.41]{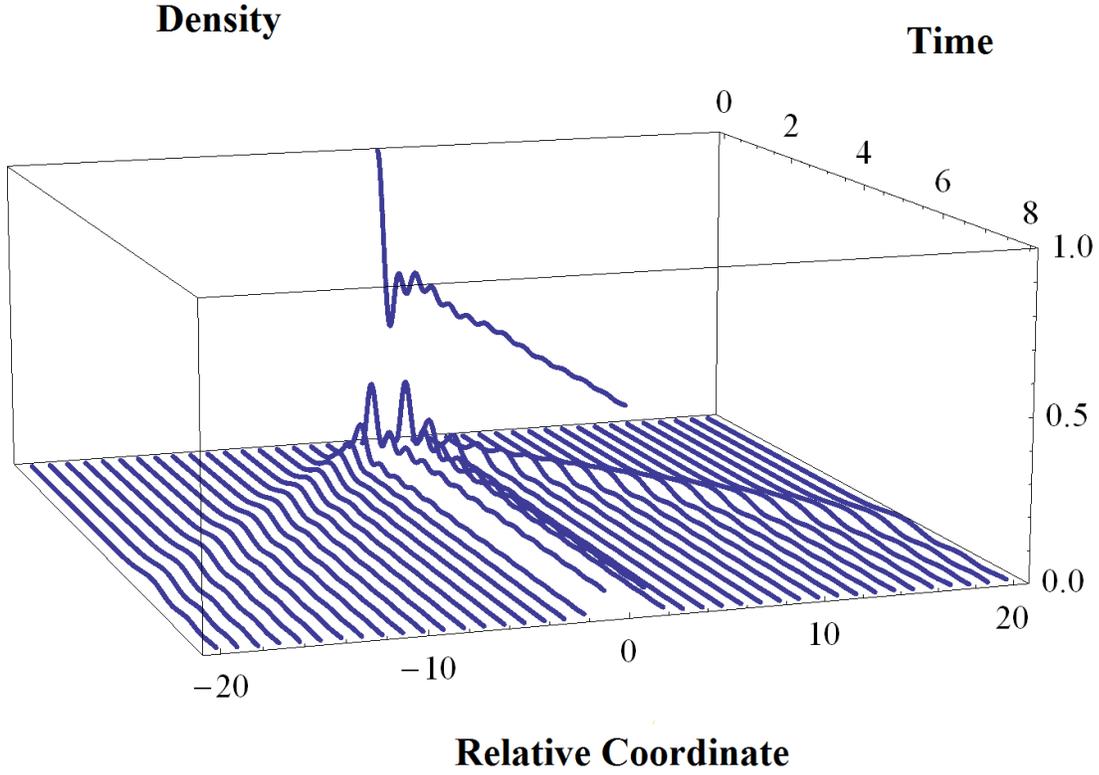}
\caption{(Color online) Density $[\rho(r,t)]$ in the relative coordinate as a function of time $(t)$ for $U=5$, $J=1$, $K=0$ and $d=0$. It can be seen that there are three wave-packets propagating in time. The one propagating to the left and the one propagating to the right (with equal amplitude due to the bosonic symmetry of the wavefunction) describe the separation of the two particles. The one with the largest amplitude is centered around the origin and describes the time-evolution of the pair. See Sect.~\ref{subsec:quantities} for more details. All quantities are dimensionless.}
\label{fig:density1a}
}
\end{figure}

\begin{figure}
\centering
{
\includegraphics[scale=0.41]{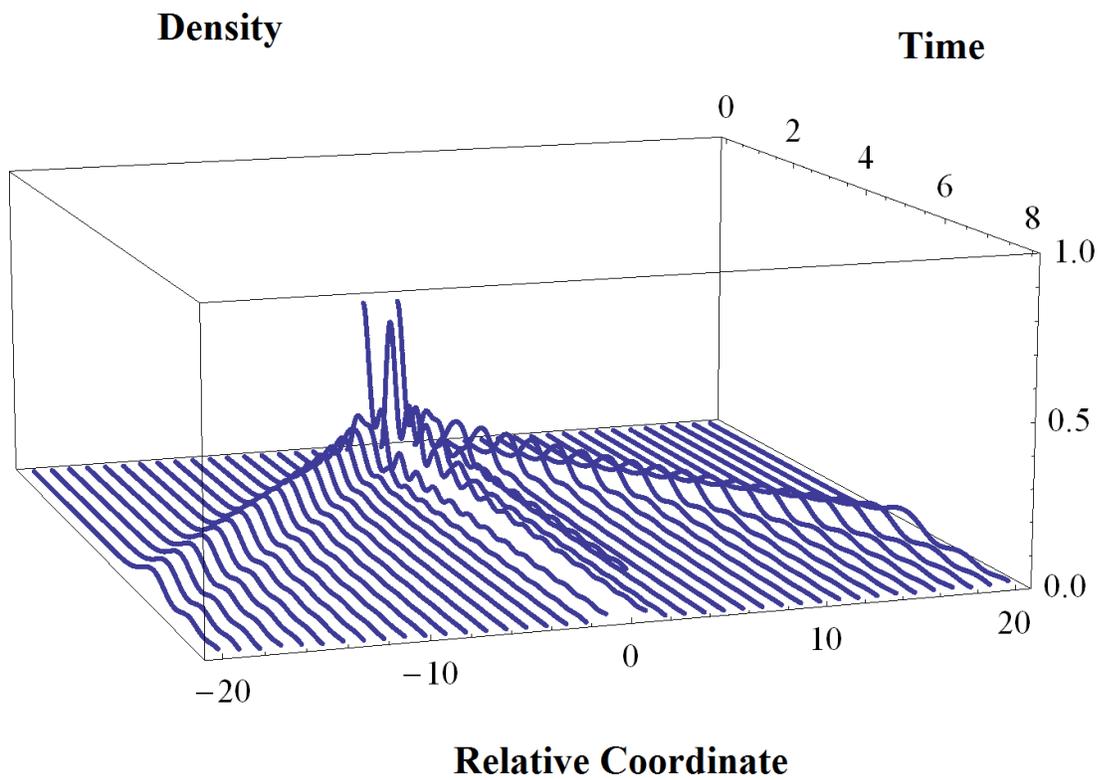}
\caption{(Color online) Same as Fig.~\ref{fig:density1a} but for $d=1$. The wave-packet in the middle is less pronounced than for $d=0$. Hence, there is a smaller probability to find a pair. See Sect.~\ref{subsec:quantities} for more details. All quantities are dimensionless.}
\label{fig:density1b}
}
\end{figure}

\begin{figure}[htb]
\centering
{
\includegraphics[scale=0.41]{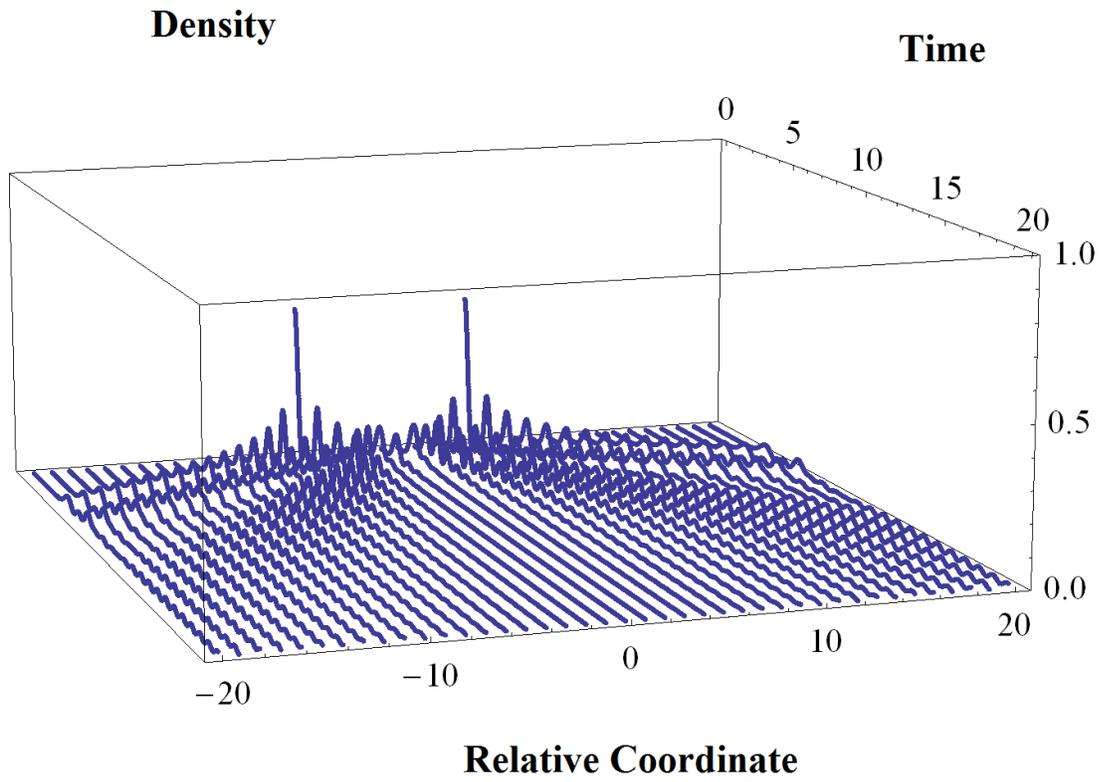}
\caption{(Color online) Same as Fig.~\ref{fig:density1a} but for $d=5$. In the dynamics with a larger initial distance new effects like a second splitting of the wave-packet appear. See Sect.~\ref{subsec:quantities} for more details. All quantities are dimensionless.}
\label{fig:density_long}
}
\end{figure}

What about the density? If we start the dynamics with two particles at one lattice site ($d=0$) and set $U=5$ the initially localized wave packet splits into three wave packets that are propagating in time, see Fig.~\ref{fig:density1a}. The wave packet centered around $r=0$ describes the time evolution of a pair while the other two wave packets describe the separation of the two particles. We remind that the density at $r=0$ equals the pair probability. The fact that we see two wave packets describing the separation of the particles is due to the bosonic symmetry of the wavefunction. For  two particles that initially sit at adjacent lattice sites ($d=1$) the picture qualitatively looks the same, see Fig.~\ref{fig:density1b}. As one would expect the wave packet describing the time evolution of the pair is less pronounced which means that the particles separate with a higher probability. If we start the dynamics with larger initial distances the dynamics of the density becomes more complicated. For $d=5$ the initial wave packet splits into four wave packages that have a much broader shape than in the case of $d=0$ or $d=1$, see Fig.~\ref{fig:density_long}. The wave packet describing the pair propagation has nearly vanished.
\section{Asymptotic dynamics}
\label{sec:asymptotic_dynamics}
In the previous section we have seen that the pair probability is converging rapidly towards an asymptotic value for large times. We are interested in this asymptotic value and therefore compute the wavefunction, the pair probability and its variance in the limit of infinite time. The calculation can easily be done with the help of the Riemann-Lebesgue lemma which is used to show that the function $I(r,t)$ defined in Eq.~\eqref{eq:dynamics7a} vanishes in the desired limit, see App.~\ref{sec:RLL}. The asymptotic wavefunction reads
\begin{align}
\Psi_{U \leq 0}^a( R, r ) &= c_d \frac{e^{i Q R}}{\sqrt{2 \pi}} \frac{ \left| U_Q \right|}{\sqrt{1+U_Q^2}} \left( \sqrt{1+U_Q^2} - \left| U_Q \right| \right)^{\left| r \right| + \left| d \right|},
\label{eq:adynamics1b} \\
\Psi_{U \geq 0}^a( R, r ) &= c_d \frac{e^{i Q R}}{\sqrt{2 \pi}} \frac{U_Q}{\sqrt{U_Q^2+1}} \left( U_Q  - \sqrt{U_Q^2+1} \right)^{\left| r \right| + \left| d \right|}.
\label{eq:adynamics1a}
\end{align} 
We note that it is not normalized anymore. This is because $\Psi(t)$ converges to $\Psi^a$ only pointwise and not in the $l^2$ norm. The same behavior can be found for example for a textbook Gaussian wavefunction describing the dynamics of a particle in free space \cite{Schwabl2002}. For all times the wavefunction is normalized. In the limit of infinite time it converges pointwise to zero but it does not converge to the constant zero function in the $L^2$ norm. In our case this property can make the calculation of other asymptotic quantities more demanding because one may not be allowed to interchange summations coming from the $l^2$ scalar product with the time limit.

\begin{figure}[htb]
\centering
{
\subfigure[]{
\includegraphics[scale=0.14]{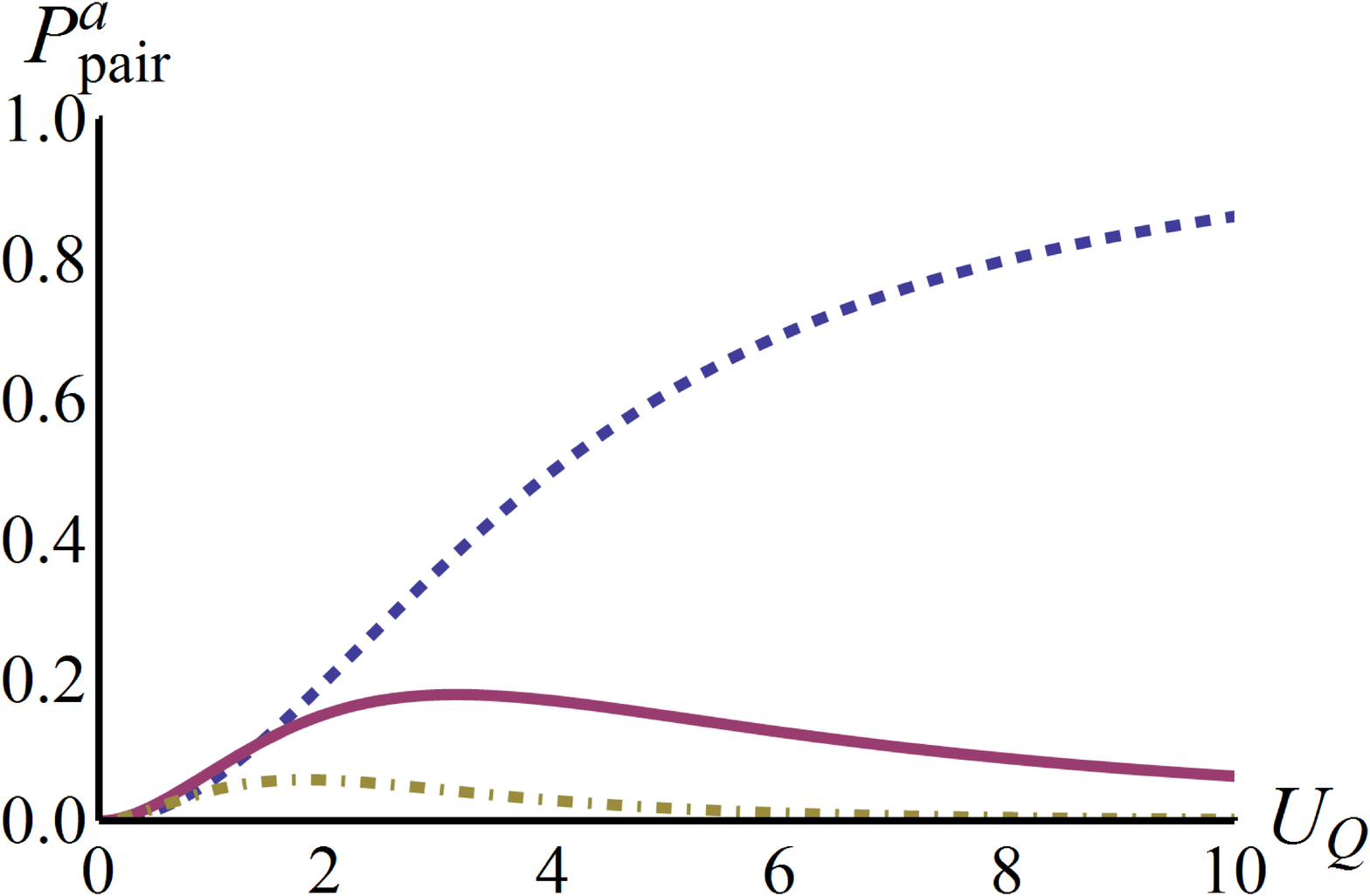}
\label{fig:adensity1a}
}
\hspace{3 mm}
\subfigure[]{
\includegraphics[scale=0.14]{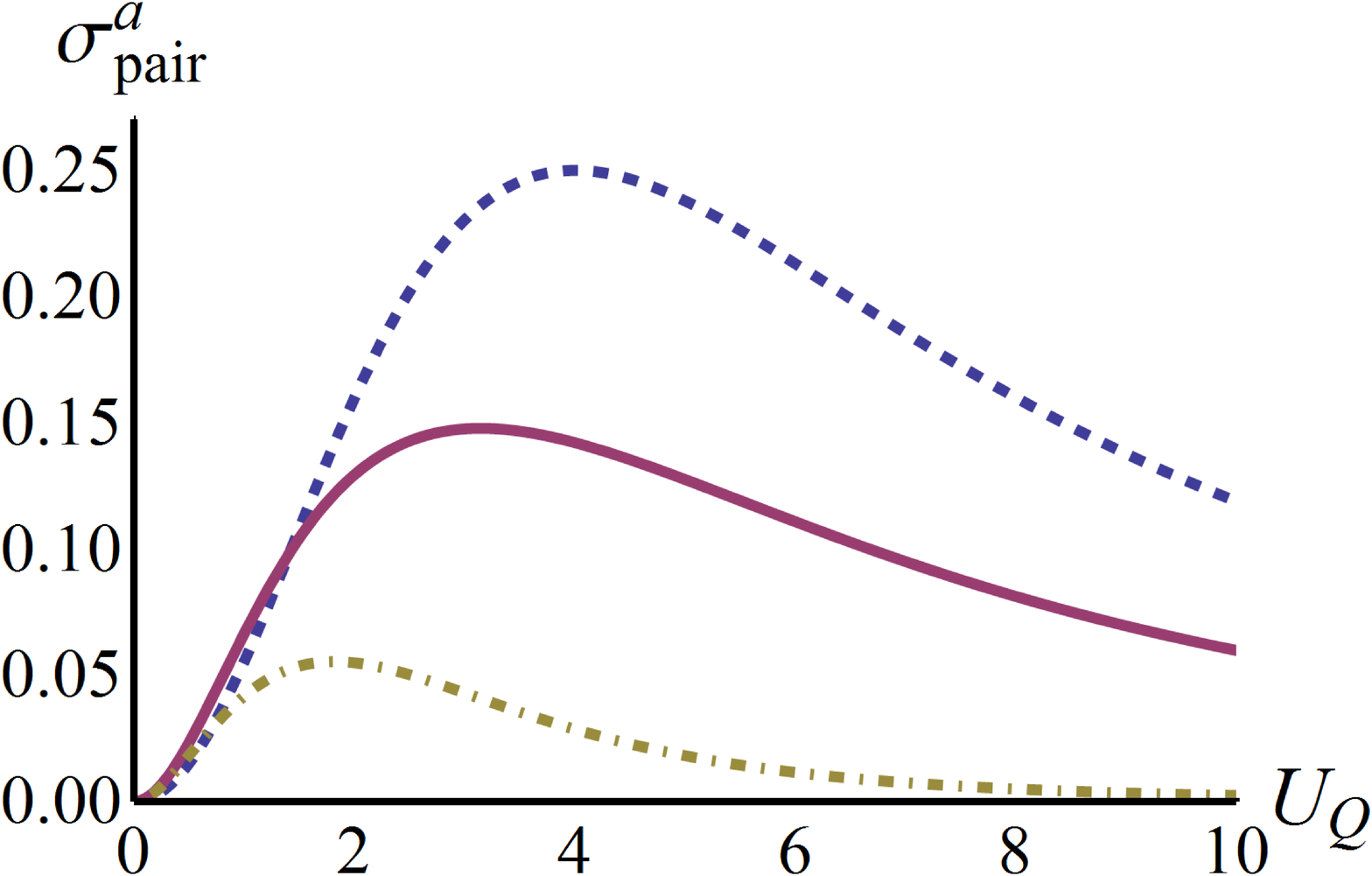}
\label{fig:adensity1b}
}
\caption{(Color online) Asymptotic pair probability $[P^a_{pair}]$ (a) and its variance $[\sigma^a_{pair}]$ (b) as a function of the effective interaction strength $U_Q$ for $J=1$ and $d=0$ [dashed blue (highest) curve], $d=1$ [solid violet (middle) curve], $d=2$ [dot-dashed brown (lowest) curve]. The asymptotic pair probability has a local maximum for $d \neq 0$, the asymptotic variance for all initial distances. Since for $d \neq 0$ the asymptotic pair probability is always smaller than $\frac{1}{2}$ its maximum and the maximum of its variance coincide, see Sect.~\ref{sec:asymptotic_dynamics} for more details. All quantities are dimensionless.}
\label{fig:adensity1}
}
\end{figure}

>From the asymptotic wavefunction we compute the asymptotic pair probability 
\begin{equation}
P_{pair}^a = c_d^2 \frac{U_Q^2}{U_Q^2+1} \left( \sqrt{1+U_Q^2} - \left| U_Q \right| \right)^{2 \left| d \right|},
\label{eq:adynamics2}
\end{equation}
where $U_Q = \frac{U}{2 J_Q} = \frac{U}{4 J \cos(Q/2)}$. For $d=0$ the formula reduces to $P_{pair}^a = \frac{U_Q^2}{U_Q^2+1}$ which is a strictly increasing function of the effective interaction parameter and approaches one for large effective interaction strengths, see Fig.~\ref{fig:adensity1a}. Hence, also from the asymptotic pair probability we can conclude that a large interaction strength or a large center of mass quasi-momentum stabilizes an initially prepared pair. As expected, the asymptotic pair probability has a local maximum for all $d \neq 0$, see Fig.~\ref{fig:adensity1a} for $d=1,2$. The location of this maximum defines the optimal interaction strength for pair formation which is given by
\begin{equation}
U_Q^{max}(d) = \pm \sqrt{ \frac{\sqrt{1 + \frac{4}{d^2}} - 1} {2} }.
\label{eq:adynamics3}
\end{equation}
We note that $U_Q^{max}(d)$ is a strictly decreasing function of the initial distance $d$. Hence, if we chose a larger initial distance the optimal effective interaction strength for pair formation will be smaller. 

As we have seen in Sect.~\ref{sec:dynamics}, the variance of the pair probability is a function of the pair probability itself. The same holds for the asymptotic variance of the pair probability which reads
\begin{equation}
\sigma_{pair}^a = P_{pair}^a \left[ 1 - P_{pair}^a \right],
\label{eq:eq:adynamics4}
\end{equation}  
see Eq.~\eqref{eq:dynpairprob2}. It has a local maximum for all values of the initial distance $d$, see Fig.~\ref{fig:adensity1b} for $d=0,1,2$. In case of $d > 0$ the asymptotic pair probability is always smaller than $\frac{1}{2}$ and we can conclude that the maxima of the asymptotic pair probability and its variance coincide. Hence, in this regime a large pair probability always leads to a large variance of this quantity. 
\section{Dynamical Symmetry in the $N$-Particle System}
\label{sec:symmetry_analysis}
In Sect.~\ref{subsec:quantities} we have shown that the time-dependent pair probability, its variance and the time-dependent density do not depend on the sign of the interaction strength $U$. This behavior is an important finding because it provides an explanation for the existence of repulsively bound atom pairs that is complementary to the one given in \cite{Repnature2006}. If a repulsive interaction leads to the same time-dependent pair probability as an attractive interaction of the same magnitude, why should there be no dynamical stability of an initially prepared pair? Nevertheless, until now we have investigated the dynamics of two particles whereas the experiment has been performed with a gas consisting of approximately $2 \cdot 10^4$ atoms. To be able to make statements also in this regime, we extend the result on the invariance of the three observables under a change of the sign of $U$ from Sect.~\ref{subsec:quantities} and show that a similar statement holds for the $N$-particle system, provided one chooses the ``right'' initial conditions. Interestingly, for its proof one does not need to specify the statistics of the particles - it works for bosons and for fermions alike. As a prerequisite we discuss simple relations between the spectra and eigenfunctions of the attractive and the repulsive Bose- or Fermi-Hubbard model.

It is well known that the Bose-Hubbard model with two lattice sites (Bose-Hubbard dimer) possesses a symmetry connecting the attractive and the repulsive Hamiltonian. This symmetry leads to relations between static properties like for example the energy spectrum of the attractive and the repulsive system \cite{Penna2001,Buonsante-2005,Links_Hibberd-2006,Links2009b}. As was first realized and quantified in \cite{PhysRevA.82.013620} the symmetry also affects the dynamics and leads to a dynamical symmetry in the Bose-Hubbard dimer. The authors could show that the Bose-Hubbard model dictates an equivalence between the time evolution of the survival probability and fragmentation of the attractive and the repulsive system if all $N$ particles initially reside in one of the two wells. A short time afterwards it was shown that the time evolution of expectation values in the Fermi-Hubbard model under certain conditions does not depend on the sign of the interaction strength \cite{breakdowndiffusion2010}. The main result we present here (Theorem~2) is an extension of some of the results to be found in \cite{PhysRevA.82.013620,breakdowndiffusion2010} for a more general class of operators and initial conditions.
\subsection{A relation between the spectra and eigenfunctions of the attractive and the repulsive model}
\label{sec:timeindependent}
We start the analysis with the definition of the unitary operator \cite{PhysRevA.82.013620}
\begin{equation}
\hat{R} = \left\{ \hat{a}_{\alpha} \rightarrow (-1)^{ \alpha } \ \hat{a}_{\alpha} \right\}.
\label{eq:symmetry1}
\end{equation}
It changes the sign of the creation and the annihilation operator at every second lattice site and thereby shifts the quasi-momentum of each particle by $\pi$, see App.~\ref{sec:roperator}. By $\hat{a}_{\alpha}$ we denote a bosonic or fermionic creation operator for a particle at lattice site $\alpha$. Possible spin indexes are surpressed since they do not play a role. For convenience we assume a one-dimensional infinite lattice $\Gamma = \mathbb{Z}$ but everything still works for a finite lattice with periodic boundary conditions for an even number of lattice sites and in higher dimensions. When we assume the Bose-Hubbard or Fermi-Hubbard Hamiltonian \footnote{The Fermi-Hubbard or just Hubbard Hamiltonian reads $\hat{H} = -J \sum_{\alpha \in \mathbb{Z}, \sigma \in \left\{ \uparrow, \downarrow \right\}} \hat{c}^{\dagger}_{\alpha,\sigma} \hat{c}_{{\alpha+1,\sigma}} + h.c. + U \sum_{\alpha \in \mathbb{Z}} \hat{n}_{\alpha, \uparrow} \hat{n}_{\alpha, \downarrow}$, where $\hat{c}^{\dagger}_{\alpha,\sigma}$/$\hat{c}_{\alpha,\sigma}$ is the fermionic creation/annihilation operator and $\hat{n}_{\alpha, \sigma}$ the number operator for a particle at site $\alpha$ with spin $\sigma$, respectively \cite{one-dimensional-hubbard}.}, the following relation holds \cite{PhysRevA.82.013620}
\begin{equation}
\hat{R} \hat{H}(U) \hat{R} = -\hat{H}(-U).
\label{eq:symmetry2}
\end{equation}
Using this equality it is easy to proof a statement about the spectrum and the eigenfunctions of the Hamiltonian $\hat{H}(U)$. \newline
\textbf{Theorem~1: }\textit{Let $\left| \Psi(U) \right\rangle$ be an eigenfunction of the Hamiltonian $\hat{H}(U)$ with eigenvalue $E(U)$. Then $\left| \Psi(-U) \right\rangle = \hat{R} \left| \Psi(U) \right\rangle$ is an eigenfunctions of $\hat{H}(-U)$ with eigenvalue $E(-U) = - E(U)$.} \newline
\textbf{Proof: } Assume we are given the solution of the time-independent Schr\"odinger equation with interaction strength $U$, $\hat{H}(U) \left| \Psi(U) \right\rangle = E(U) \left| \Psi(U) \right\rangle$. When we let $\hat{R}$ act on both sides of this equation we find $\hat{H}(-U) \hat{R} \left| \Psi(U) \right\rangle = - E(U) \hat{R} \left| \Psi(U) \right\rangle$. $\blacksquare$

Hence, the simple relation between the Hamiltonian of the attractive and the repulsive model leads as well to simple relations between their spectra and eigenfunctions. Theorem~1 can also be interpreted as a statement about the existence of repulsively bound states in the two models which can be seen as follows. We know that the wavefunctions of the attractive and the repulsive system are related by the operator $\hat{R}$. Translated to first quantization this relation reads $\Psi(x_1,...,x_N,-U) = (-1)^{x_1+...+x_N} \Psi(x_1,...,x_N,U)$ (see App.~\ref{sec:roperator}), and consequently the density of $\Psi(U)$ and of $\Psi(-U)$ equal each other. From this it follows that if $\Psi(U)$ is square summable the same must be true for $\Psi(-U)$. Therefore, the Bose-Hubbard and the Fermi-Hubbard model with attractive and repulsive interaction of the same magnitude have an equal number of bound states (square summable wavefunction). This surprising finding is a pure lattice effect and contrasts with quantum mechanics in a continuous coordinate space where potentials usually do not have bound states anymore when their character is changed from attractive to repulsive. 

As an example we mention the solution of the time-independent Schr\"odinger equation [Eq.~\eqref{eq:dynamics3}] from Sect.~\ref{subsec:wavefunction}. As already mentioned the spectrum of the Hamiltonian consists of a scattering continuum [Eq.~\eqref{eq:dynamics5}] whose energy is invariant under the operation $E \rightarrow -E$ and a bound state below or above the scattering continuum for attractive and repulsive interaction, respectively [Eq.~\eqref{eq:dynamics4}]. The bound state wavefunctions of the two systems differ by a factor of $(-1)^{\left| r \right|} = (-1)^{x + y}$. 
\subsection{Invariance of time-dependent expectation values under the transformation~$U~\rightarrow~-U$}
\label{sec:timedependent}
As we have seen in the previous subsection, Eq.~\eqref{eq:symmetry1} leads to simple relations between the spectra and eigenfunctions of the attractive and the repulsive system. This certainly affects also the dynamics. Upon operation of $\hat{R}$, the transformation of the spectrum ($E \rightarrow -E$) leads to a reversal of time; the transformation of the wavefunction ($\Psi \rightarrow \hat{R} \Psi$) is not that intuitive, see App.~\ref{sec:roperator}. Nevertheless, with a little more effort needed to derive the time-independent result it is possible to make a statement about time-dependent properties as well. We start with a definition. \newline 
\textbf{Definition: }\textit{We call an operator $\hat{O}$ real if it has only real coefficients when being expressed with the creation and annihilation operators $\hat{a}_{\alpha}$ and $\hat{a}_{\alpha}^{\dagger}$. An operator that is given by matrix elements in the Wannier basis  \footnote{\label{footnote1} We use the term Wannier basis in the sense of \cite{one-dimensional-hubbard}. The Wannier basis or multi-particle Wannier basis ($N>1$) is given by symmetrized (permanents) or antisymmetrized (determinants) tensor products of Kronecker deltas $\Psi_{\alpha}(x) = \delta_{\alpha,x}$ which describe localized bosons or fermions (at point $\alpha \in \mathbb{Z}$), respectively. This should not be confused with Wannier functions which are introduced in continuous space \cite{Wannier1937}.} 
is called real if all its matrix elements are real.} \newline
Using the definition we state the main result of this section: \newline
\textbf{Theorem~2: }\textit{Let $\hat{O}$ be a hermitian operator that can be written as $\hat{O} = \hat{O}_{r} + i \hat{O}_i$, where $\hat{O}_{r}$ and $\hat{O}_i$ are real operators and fulfill the relations $\hat{R} \ \hat{O}_{r} \ \hat{R} = \hat{O}_{r}$ and $\hat{R} \ \hat{O}_{i} \ \hat{R} = - \hat{O}_{i}$. We assume that $\hat{H}(U)$ is the Bose-Hubbard or Fermi-Hubbard Hamiltonian. The initial condition reads $\Psi_0  = \varphi + i \chi$ where $\varphi$ and $\chi$ are assumed to be real functions on the lattice $\Gamma^N = \mathbb{Z}^N$. Additionally, $\varphi$ and $\chi$ are eigenfunctions of $\hat{R}$ with different eigenvalues, in formulas $\hat{R} \left| \varphi \right\rangle = \pm \left| \varphi \right\rangle$ and $\hat{R} \left| \chi \right\rangle = \mp \left| \chi \right\rangle$. Then the following relation for the time-dependent expectation value of $\hat{O}$ is true: $ O(U,t) = O(-U,t)$.} \newline
\textbf{Proof:} We give here the proof only for the special case when $\Psi_0$ is a real function on $\Gamma^N$ and $\hat{O}$ is a real operator in analogy to \cite{PhysRevA.82.013620}. The general proof can be found in App.~\ref{sec:prooftheorem2}. We write the expectation value of $\hat{O}$ as
\begin{align}
O(U,t) &= \left\langle \Psi_0 \right| e^{i \hat{H}_{U} t} \hat{O} \ e^{-i \hat{H}_{U} t} \left| \Psi_0 \right\rangle 
\label{eq:symmetry3} \\
	&= \left\langle \Psi_0 \right| \cos(\hat{H}_{U} t) \hat{O} \cos(\hat{H}_{U} t) \left| \Psi_0 \right\rangle \nonumber \\
	&\ \ \ + \left\langle \Psi_0 \right| \sin(\hat{H}_{U} t) \hat{O} \sin(\hat{H}_{U} t) \left| \Psi_0 \right\rangle \nonumber \\
	&\ \ \ - 2 \operatorname{Im} \left[ \left\langle \Psi_0 \right| \sin(\hat{H}_{U} t) \hat{O} \cos(\hat{H}_{U} t) \left| \Psi_0 \right\rangle \right]. \nonumber
\label{eq:symmetry3}
\end{align}
The problem we have to manage is to change the sign of $U$ with an insertion of $\hat{R}^2 = \mathds{1}$ factors without changing the direction of time. This is because $\hat{R} e^{-i \hat{H}_{U} t} \hat{R} = e^{i \hat{H}_{-U} t}$. On that account, let us analyze the matrix element $\left\langle \Psi_0 \right| \sin(\hat{H}_{U} t) \hat{O} \cos(\hat{H}_{U} t) \left| \Psi_0 \right\rangle$. All operators have only real coefficients when being expressed with $\hat{a}_{\alpha}$ and $\hat{a}_{\alpha}^{\dagger}$ or when being expanded in the Wannier basis. Additionally, the creation and annihilation operators produce only real numbers when acting on occupation number states (or Wannier basis states which is the same in this setting) and the wave function at time zero $\Psi_0$ can be expanded into the Wannier basis with real coefficients only (it is a real function on $\Gamma^N$). Hence, the overall matrix element is real and its contribution to the expectation value of $\hat{O}$ vanishes. We find
\begin{equation}
O(U,t) = \left\langle \Psi_0 \right| \cos(\hat{H}_{U} t) \hat{O} \cos(\hat{H}_{U} t) + \sin(\hat{H}_{U} t) \hat{O} \sin(\hat{H}_{U} t) \left| \Psi_0 \right\rangle.
\label{eq:symmetry4}
\end{equation}
Now we can insert $\hat{R}^2 = \mathds{1}$ factors between all operators and thereby change the sign of the interaction strength $U$ without changing the direction of time. When we additionally use the two relations $\hat{R} \hat{O} \hat{R}=\hat{O}$ and $\hat{R} \Psi_0 = \pm \Psi_0$ the result of Theorem~2 can be shown. $\blacksquare$

Let us discuss the applications of Theorem~2. First we note that it applies to the two-particle dynamics of Sect.~\ref{sec:dynamics}. The initial condition $\Psi_{0,0}(R,r) = \frac{e^{i Q R}}{\sqrt{2 \pi}} \delta_{r,0}$ reads $\left| \Psi_{0,0} \right\rangle = \sum_{x,y \in \Gamma} \frac{e^{i Q (x+y)/2}}{\sqrt{2 \pi}} \delta_{x,y} \hat{b}_x^{\dagger} \hat{b}_y^{\dagger} \left| 0 \right\rangle = \sum_{x \in \Gamma} \frac{e^{i Q x}}{\sqrt{2 \pi}} \left( \hat{b}_x^{\dagger} \right)^2 \left| 0 \right\rangle$ in second quantization, also see App.~\ref{sec:roperator}. When we let $\hat{R}$ act on this state we find $\hat{R} \left| \Psi_0 \right\rangle = \left| \Psi_0 \right\rangle$, but obviously the expansion coefficients are not real. This can be circumvented by using the Hamiltonian of Eq.~\eqref{eq:dynamics3} as the starting point. In second quantization it reads $\hat{H}(U) = - J_K \sum_{\alpha \in \Gamma} \hat{b}_{\alpha}^{\dagger} \hat{b}_{\alpha+1} + h.c. + U \hat{n}_0$ and fulfills Eq.~\eqref{eq:symmetry2} as well. Restricted to the one-particle subspace (of the Fock space) its dynamics are the one governed by the original Hamiltonian in the relative coordinate (the wavefunction in the center of mass coordinate is constant in time), and hence as initial condition we have to use $\Psi^r_{0,0}(r) = \delta_{r,0}$. In second quantization this reads $\left| \Psi^r_{0,0} \right\rangle = \hat{b}_{0}^{\dagger} \left| 0 \right\rangle$. It still fulfills the relation $\hat{R} \left| \Psi_0 \right\rangle = \left| \Psi_0 \right\rangle$, and additionally has real expansion coefficients. The same calculation can be done with the initial condition $\Psi^r_{0,d}(r) = \frac{1}{\sqrt{2}} \left( \delta_{r,d} + \delta_{r,-d} \right)$ which in second quantization reads $\left| \Psi^r_{0,d} \right\rangle = \frac{1}{\sqrt{2}} \left( \hat{b}_{d}^{\dagger} + \hat{b}_{-d}^{\dagger} \right) \left| 0 \right\rangle$. Acting with $\hat{R}$ on the state we find $\hat{R} \left| \Psi_{0,d} \right\rangle = (-1)^d \left| \Psi_{0,d} \right\rangle$. Hence, $\Psi^r_{0,0}$ and $\Psi^r_{0,d}$ qualify as possible initial conditions. As observables we used the time-dependent pair probability, its variance and the time-dependent density. For their computation we have to evaluate expectation values of the operators $\hat{P}_{pair} = \sum_{\alpha \in \Gamma} \left| n_{\alpha} = 2 \right\rangle \left\langle n_{\alpha} = 2 \right|$ which, when acting on one-particle states in the relative coordinate, can be written as $\hat{P}_{pair} = \hat{b}_{r=0}^{\dagger} \hat{b}_{r=0}$ and $\hat{n}_r = \hat{b}_{r}^{\dagger} \hat{b}_{r}$. Obviously, $\hat{P}_{pair}$ and $\hat{n}_r$ both are real and fulfill the relations $\hat{R} \hat{P}_{pair} \hat{R} = \hat{P}_{pair}$ and $\hat{R} \hat{n}_r \hat{R} = \hat{n}_r$. Therefore, all three observables and with a little trick also the initial conditions qualify for an application of Theorem~2. The result explains why the dynamics of our observables does not depend on the sign of the interaction strength $U$.
 
How can Theorem~2 be used to extend the invariance properties of the pair probability, its variance and of the density to the $N$-particle system? First, we need to find appropriate observables. The pair probability (probability to find two particles at the same lattice site) for $N$ particles can be written as the expectation value of a pair operator as well. It reads \cite{PhysRevA.81.045601}
\begin{equation}
\hat{P}_{pair} = \frac{2}{N} \sum_{n \in \mathbb{N}} \left| \varphi_n \right\rangle m_n \left\langle \varphi_n \right|,
\label{eq:symmetry5}
\end{equation}
where by $\left\{ \varphi_n \right\}_{n \in \mathbb{N}}$ we denote the Wannier basis of the $N$-particle Hilbert space \footnotemark[\value{footnote}]. The number $m_n$ counts the pairs in each Wannier basis state and the prefactor $\frac{2}{N}$ assures the normalization. We highlight that the pair operator is a $N$-particle operator, and consequently the full wavefunction is needed to compute its expectation value. The fact that the Wannier basis states are eigenfunctions of $\hat{R}$ \footnote{In the occupation number representation a Wannier basis state reads $\left| n_{\alpha_1}, n_{\alpha_2}, ..., n_{\alpha_p} \right\rangle = \frac{1}{\sqrt{n_{\alpha_1} ! n_{\alpha_2} ! ... n_{\alpha_p} !}} \left( \hat{a}_{\alpha_1}^{\dagger} \right)^{n_{\alpha_1}} \cdot \left( \hat{a}_{\alpha_2}^{\dagger} \right)^{n_{\alpha_2}} \cdot ... \cdot \left( \hat{a}_{\alpha_p}^{\dagger} \right)^{n_{\alpha_p}} \left| 0 \right\rangle$ where possible spin indexes have been suppressed. The numbers $n_i$ ($i=1,...,p$) count the particles in the state $\left| n_{\alpha_1}, n_{\alpha_2}, ..., n_{\alpha_p} \right\rangle$ that are described by the one-particle wavefunction $\Psi_{\alpha_i}(x) = \delta_{\alpha_i,x}$. Acting with $\hat{R}$ on the state we find $\hat{R} \left| n_{\alpha_1}, n_{\alpha_2}, ..., n_{\alpha_p} \right\rangle = (-1)^{n_{\alpha_1} + n_{\alpha_2} + ... + n_{\alpha_p}} \left| n_{\alpha_1}, n_{\alpha_2}, ..., n_{\alpha_p} \right\rangle$. We note that in case of fermionic particles the single occupation numbers are restricted to be equal to zero or one.} can be used to show that $\hat{R} \ \hat{P}_{pair} \ \hat{R} = \hat{P}_{pair}$ holds. For the computation of the variance of the pair probability we also need to compute the expectation value of $\hat{P}_{pair}^2$ which has the same properties. As an equivalent for the density in the two-particle system we choose the $p$-particle density ($p \leq N$) that is given by the expectation value of the operator
\begin{equation}
\hat{n}^p(x_1,...,x_p) = \frac{(N-p)!}{N!} \ \hat{a}^{\dagger}_{x_1} \cdot ... \cdot \hat{a}^{\dagger}_{x_p} \hat{a}_{x_p} \cdot ... \cdot \hat{a}_{x_1}.
\label{eq:symmetry6}
\end{equation}
Obviously, it qualifies for an application of Theorem~2 as well. As initial states for the two-particle dynamics we have chosen particles that (in the relative coordinate) are described by Wannier basis states, and on account of this describe particles that are localized to single lattice sites. Since Wannier basis states describe real functions on $\Gamma^N$ and are eigenfunctions of $\hat{R}$ \footnotemark[\value{footnote}], we can choose them as possible initial conditions for the $N$-particle dynamics as well. Nevertheless, Theorem~2 tells us that the class of all possible initial states is much larger.

We conclude that Theorem~2 generalizes the invariance properties of the pair probability, its variance and the density under the transformation $U \rightarrow - U$ to a system with $N$ bosonic or fermionic particles. Additionally, Theorem~1 tells us that repulsively bound states are not a speciality of the two-particle system but will always occur when the attractive system has bound states. The experiment on repulsively bound atom pairs has been performed with atom pairs that initially have been localized to single lattice sites. Whether these initial states qualify for an application of Theorem~2 is difficult to say. Nevertheless, if we approximate the experimental initial state with a pure Wannier basis state we can show that the dynamics do not depend on the sign of the interaction strength $U$. The stability of repulsively interacting pairs therefore becomes very intuitive when we understand that the lattice structure of the coordinate space forces the attractive and the repulsive system to act very similar.  
\section{Summary and Outlook}
\label{sec:summary}
To summarize the findings, we have computed an exact expression for the time-dependent wavefunction for two bosons trapped in an infinite one-dimensional optical lattice potential within the framework of the Bose-Hubbard model. As initial conditions we have chosen localized atoms that are separated by a distance of $d$ lattice sites and carry a center of mass quasi-momentum. An initially localized pair
($d = 0$) is found to be more stable as quantified by the pair probability when the interaction and/or the center of mass quasi-momentum is increased. In contrast, for two initially separated atoms there exists an optimal interaction strength for pair formation. 

To gain further information we have monitored the variance of the pair probability and the density in the relative coordinate during the dynamical process. Analytical expressions for the wavefunction, the pair probability and the optimal interaction strength for pair formation have been derived in the limit of infinite time. We had to give two distinct expressions for the time-dependent wavefunction for positive and negative interaction strength. In contrast, the pair probability, its variance and the density in the relative coordinate are invariant under the transformation $U \rightarrow -U$. This leads to the conclusion that for our initial conditions the three observables have the same dynamics when being propagated with the attractive or with the repulsive Hamiltonian.

In the second part of the paper we have extended this result and shown that also in the $N$-particle system there exist time-dependent observables that stay the same when the sign of the interaction strength $U$ is changed. The time-dependent pair probability of the $N$-particle system, its variance and the time-dependent $p$-particle density ($p \leq N$) belong to this class. Additionally, we have discussed a simple relation between the spectra and eigenfunctions of the attractive and the repulsive Bose- or Fermi-Hubbard model. By showing that the dynamics of the pair probability is the same in the attractive and in the repulsive $N$-particle system, we provide a complementary understanding for the recently observed \cite{Repnature2006} dynamical stability of atom pairs in a repulsively interacting lattice gas.  

The explicit expression for the time-dependent wavefunction we have computed allows one for studies which go far beyond the scope of this work. It would be an interesting and challenging task to compute for example the one-particle reduced density matrix and to study questions of entanglement during the dynamical process. Additionally, there is evidence that the computation of the exact time-dependent wavefunction on the infinite Bose-Hubbard lattice is possible for other classes of initial conditions, too. 

The finding that the Bose- or Fermi-Hubbard Hamiltonian with attractive and repulsive interaction have an equal number of bound states suggests that there is hope to experimentally find also repulsively bound states consisting of three \cite{PhysRevA.81.011601} or possibly even more particles. In addition, the result on the equivalence of expectation values under the transformation $U \rightarrow -U$ can be applied to the probability to find $m$ ($m \leq N$) particles at one lattice site as well. The same holds for the probability to find two particles at one lattice site and another one at a neighboring lattice site, which would model the situation of a bound state consisting of a dimer and a monomer. The latter suggests that even though such objects lie energetically above the scattering continuum, they nevertheless should be dynamically stable when being part of a dilute lattice gas.

Financial support by the DFG is gratefully acknowledged.
\begin{appendix}
\section{Asymptotic behavior of $I(r,t)$}
\label{sec:RLL}
To compute the asymptotic behavior of the wave function, the pair probability and its variance, we need to compute the infinite time limit of the function $I(r,t)$, see Eq.~\eqref{eq:dynamics7a}. This can be done with the help the Riemann-Lebesgue Lemma which we state in the following form \cite{Ledermann1982}: \newline
\textit{Riemann-Lebesgue Lemma}: Let $f$ be Lebesgue-integrable on $\left[ -\pi, \pi \right]$, then
\begin{equation}
\lim_{p \rightarrow \infty} \int_{-\pi}^{\pi} f(x) \ e^{i p x} dx = 0.
\label{eq:RL1}
\end{equation}
In order to bring $I(r,t)$ to a form that the Lemma is applicable, we do the coordinate transformation $q = \cos(k)$. Using the relation $\sin[\arccos(q)] = \sqrt{1-q^2}$ we find
\begin{align}
I(r,t) &= \int_{-1}^{1} \frac{dq}{\pi} \frac{e^{i 2 J_Q t q}}{\sqrt{1-q^2} + \frac{U_Q^2}{\sqrt{1-q^2}}} f_d[\arccos(q)] f_r[\arccos(q)] 
\label{eq:RL2}
\end{align}
with $f_n(k) = \cos(k n) + \frac{U_Q}{\sin(k)} \sin(k \left| n \right|)$. We note that the function $f_n[\arccos(q)]$ is continuous for all $n \in \mathbb{Z}$. The expression $\frac{1}{ \sqrt{1-q^2} + \frac{U_Q^2}{\sqrt{1-q^2}} }$ in the denominator of Eq.~\eqref{eq:RL2} is continuous as well, and hence the Riemann-Lebesgue Lemma is applicable, leading to the result  
\begin{equation}
\lim_{t \rightarrow \infty} I(r,t) = 0.
\label{eq:RL3}
\end{equation}
Therefore, the contribution from the scattering states to the pair probability and to the density vanishes in the limit of infinite time.
\section{The action of the operator $\hat{R}$}
\label{sec:roperator}
In this appendix we provide formulas for the action of the operator $\hat{R}$ defined in Eq.~\eqref{eq:symmetry1} on the annihilation operator of a particle with quasi-momentum $k$ and on the wavefunction in first quantization in coordinate and in quasi-momentum space. Since the wavefunction in second quantization can be expressed with creation and annihilation operators, the action of $\hat{R}$ on it is obvious. To compute its action on the wavefunction in first quantization we recall the relation between first and second quantized wavefunctions
\begin{equation}
\left| \Psi \right\rangle = \sum_{x_1,...,x_N \in \Gamma} \Psi(x_1,...,x_N) \ \hat{a}^{\dagger}_{x_1} ... \hat{a}^{\dagger}_{x_N} \left| 0 \right\rangle.
\label{eq:appRoperator1}
\end{equation}
Here $\left| \Psi \right\rangle$ denotes the wavefunction in second quantization, $\Psi(x_1,...,x_N)$ the wavefunction in first quantization and $\left| 0 \right\rangle$ is the vacuum state. As already defined in Sect.~\ref{sec:symmetry_analysis} the operator $\hat{a}^{\dagger}_{x}$ denotes a bosonic or fermionic creation operator for a particle at site $x \in \Gamma$. Possible spin indexes are suppressed. When acting with $\hat{R}$ on $\left| \Psi \right\rangle$ we find $\hat{R} \left| \Psi \right\rangle = \sum_{x_1,...,x_N \in \Gamma} \Psi(x_1,...,x_N) (-1)^{x_1 + ... + x_N} \ \hat{a}^{\dagger}_{x_1} ... \hat{a}^{\dagger}_{x_N} \left| 0 \right\rangle$, and hence the action of $\hat{R}$ on the first quantized wavefunction is given by
\begin{equation}
\hat{R} \Psi(x_1,...,x_N) = (-1)^{x_1 + ... + x_N} \Psi(x_1,...,x_N).
\label{eq:appRoperator2}
\end{equation}
As one could expect $\hat{R}$ changes its sign at every second lattice site. For two (or more) particles this happens in a chess-pattern-like way. 

How does $\hat{R}$ act on functions in momentum space? The annihilation operator of a particle with quasi-momentum $k$ reads $\hat{a}_k = \frac{1}{\sqrt{2 \pi}} \sum_{x \in \Gamma} e^{i k x} \hat{a}_x$. Using the relation $\hat{R} \hat{a}_x \hat{R} = (-1)^x \hat{a}_x$ we find
\begin{equation}
\hat{R} \hat{a}_k \hat{R} = \hat{a}_{k + \pi}.
\label{eq:appRoperator3}
\end{equation} 
Having Eq.~\eqref{eq:appRoperator3} at hand we can compute the action of $\hat{R}$ on the first quantized wavefunction in quasi-momentum space $\tilde{\Psi}(k_1,...,k_N)$. When we write the second quantized wavefunction as $\left| \Psi \right\rangle = \frac{1}{\sqrt{2 \pi}^N} \int_{-\pi}^{\pi} d k_1 ... \int_{-\pi}^{\pi} d k_N \tilde{\Psi}(k_1,...,k_N) \ \hat{a}^{\dagger}_{k_1} ... \hat{a}^{\dagger}_{k_N} \left| 0 \right\rangle$ and act with $\hat{R}$ on it we find
\begin{equation}
\hat{R} \tilde{\Psi}(k_1,...,k_N) = \tilde{\Psi}(k_1 + \pi,...,k_N + \pi).
\label{eq:appRoperator4}
\end{equation}
Hence, the operator $\hat{R}$ shifts the quasi-momentum of each particle by an amount of $+ \pi$. 
\section{Proof of Theorem~2}
\label{sec:prooftheorem2}
In Sect.~\ref{sec:timedependent} we have given the proof of Theorem~2 only for the special case when $\Psi_0$ is a real function on $\Gamma^N$ and $\hat{O}$ is a real operator. Here we assume the general scenario of Theorem~2. The initial condition reads $\Psi_0 = \varphi + i \chi$, where $\varphi$ and $\chi$ are real functions on $\Gamma^N$ and eigenfunctions of $\hat{R}$ with different eigenvalues, in formulas $\hat{R} \left| \varphi \right\rangle = \pm \left| \varphi \right\rangle$, $\hat{R} \left| \chi \right\rangle = \mp \left| \chi \right\rangle$. The observable reads $\hat{O} = \hat{O}_r + i \hat{O}_i$ with real operators $\hat{O}_{r}$ and $\hat{O}_i$. Additionally, we assume that they fulfill the relations $\hat{R} \ \hat{O}_{r} \ \hat{R} = \hat{O}_{r}$ and $\hat{R} \ \hat{O}_{i} \ \hat{R} = - \hat{O}_{i}$. The expectation value of $\hat{O}(t) = e^{i \hat{H}_U t} \hat{O} e^{-i \hat{H}_U t}$ then reads
\begin{equation}
O(U,t) = \left\langle \varphi \right| \hat{O}(t) \left| \varphi \right\rangle + \left\langle \chi \right| \hat{O}(t) \left| \chi \right\rangle - 2 \operatorname{Im} \left[ \left\langle \varphi \right| \hat{O}(t) \left| \chi \right\rangle \right].  
\label{eq:prooftheorem21}
\end{equation}
Let us have a look at the first term of Eq.~\eqref{eq:prooftheorem21}. It reads $\left\langle \varphi \right| \hat{O}(t) \left| \varphi \right\rangle = \left\langle \varphi \right| \hat{O}_r(t) \left| \varphi \right\rangle + \left\langle \varphi \right| i \hat{O}_i(t) \left| \varphi \right\rangle$. The term $\left\langle \varphi \right| \hat{O}_r(t) \left| \varphi \right\rangle$ is invariant under a change of the sign of the interaction strength $U$ because $\varphi$ is a real function and $\hat{O}_r$ is a real operator, see Sect.~\ref{sec:timedependent}. The other term can be written as
\begin{align}
\left\langle \varphi \right| i \hat{O}_i(t) \left| \varphi \right\rangle &= i \left\langle \varphi \right| \cos(\hat{H}_U t) \hat{O}_i \cos(\hat{H}_U t) + \sin(\hat{H}_U t) \hat{O}_i \sin(\hat{H}_U t) \left| \varphi \right\rangle \label{eq:prooftheorem22} \\ 
																																				 &\ \ \ - 2 \operatorname{Re} \left[ \left\langle \varphi \right| \sin(\hat{H}_U t) \hat{O}_i \cos(\hat{H}_U t) \left| \varphi \right\rangle \right] \nonumber.
\end{align}
Since $\hat{O}_i$ is a real operator the first matrix element is real. This is because all operators have only real coefficients when being expressed with $\hat{a}_{\alpha}$ and $\hat{a}_{\alpha}^{\dagger}$ or when being expanded in the Wannier basis. The creation and annihilation operators produce only real numbers when acting on occupation number states (or Wannier basis states which is the same in this setting) and $\left| \varphi \right\rangle$ can be expanded into the Wannier basis with real coefficients only (it is a real function on $\Gamma^N$). The matrix element is multiplied by the imaginary unit $i$, and hence its contribution to the expectation value $O(U,t)$ is purely imaginary. Because $O(U,t)$ is a real function this contribution has to vanish, thus $\left\langle \varphi \right| \cos(\hat{H}_U t) \hat{O}_i \cos(\hat{H}_U t) + \sin(\hat{H}_U t) \hat{O}_i \sin(\hat{H}_U t) \left| \varphi \right\rangle = 0$. The matrix element in the last term of Eq.~\eqref{eq:prooftheorem22} is real because of the same reasons which leads to
\begin{equation}
\left\langle \varphi \right| i \hat{O}_i(t) \left| \varphi \right\rangle = - 2 \left\langle \varphi \right| \sin(\hat{H}_U t) \hat{O}_i \cos(\hat{H}_U t) \left| \varphi \right\rangle.
\label{eq:prooftheorem23}
\end{equation} 
Now we can again insert $\hat{R}^2 = \mathds{1}$ factors between all operators and find 
\begin{equation}
\left\langle \varphi \right| i \hat{O}_i(t) \left| \varphi \right\rangle = - 2 \left\langle \varphi \right| \hat{R} \sin(\hat{H}_{-U} t) \hat{R} \hat{O}_i \hat{R} \cos(\hat{H}_{-U} t) \hat{R} \left| \varphi \right\rangle.
\label{eq:}
\end{equation}
>From $\hat{R} \sin(\hat{H}_U t) \hat{R} = - \sin(\hat{H}_{-U} t)$ we get a factor of $(-1)$ which cancels with the one coming from $\hat{R} \hat{O}_i \hat{R} = - \hat{O}_i$. Similar analysis holds for the second term in Eq.~\eqref{eq:prooftheorem21}. We conclude that $\left\langle \varphi \right| \hat{O}(t) \left| \varphi \right\rangle$ and $\left\langle \chi \right| \hat{O}(t) \left| \chi \right\rangle$ have the wanted invariance property. 

What remains to check is the last term of Eq.~\eqref{eq:prooftheorem21}. It reads
\begin{equation}
\operatorname{Im} \left[ \left\langle \varphi \right| \hat{O}(t) \left| \chi \right\rangle \right] = \operatorname{Im} \left[ \left\langle \varphi \right| \hat{O}_r(t) \left| \chi \right\rangle \right] + \operatorname{Im} \left[ \left\langle \varphi \right| i \hat{O}_i(t) \left| \chi \right\rangle \right].
\label{eq:prooftheorem24}
\end{equation}
The first term of the right hand side of this equation can be written as
\begin{align}
\operatorname{Im} \left[ \left\langle \varphi \right| \hat{O}_r(t) \left| \chi \right\rangle \right] &= \operatorname{Im} \left[ \left\langle \varphi \right| \cos(\hat{H}_U t) \hat{O}_r \cos(\hat{H}_U t) + \sin(\hat{H}_U t) \hat{O}_r \sin(\hat{H}_U t) \left| \chi \right\rangle \right. \label{eq:prooftheorem25} \\ 
																																				                             &\ \ \ \left. + i \left\langle \varphi \right| \sin(\hat{H}_U t) \hat{O}_r \cos(\hat{H}_U t) - \cos(\hat{H}_U t) \hat{O}_r \sin(\hat{H}_U t) \left| \chi \right\rangle \right] \nonumber \\
																		                                                                 &= \left\langle \varphi \right| \sin(\hat{H}_U t) \hat{O}_r \cos(\hat{H}_U t) - \cos(\hat{H}_U t) \hat{O}_r \sin(\hat{H}_U t) \left| \chi \right\rangle \nonumber.
\end{align}
To derive the result we have used the same argumentation as above to show that the matrix elements are real. We insert the $\hat{R}^2 = \mathds{1}$ factors to change the sign of $U$ and thereby produce a factor of $(-1)$ coming from $\hat{R} \sin(\hat{H}_U t) \hat{R} = -\sin(\hat{H}_{-U} t)$ which cancels with another one coming either from $\hat{R} \left| \varphi \right\rangle = \pm \left| \varphi \right\rangle$ or from $\hat{R} \left| \chi \right\rangle = \mp \left| \chi \right\rangle$. Using the same arguments, the second term of Eq.~\eqref{eq:prooftheorem24} can be written as
\begin{equation}
\operatorname{Im} \left[ \left\langle \varphi \right| i \hat{O}_i(t) \left| \chi \right\rangle \right] = \left\langle \varphi \right| \cos(\hat{H}_U t) \hat{O}_i \cos(\hat{H}_U t) + \sin(\hat{H}_U t) \hat{O}_i \sin(\hat{H}_U t) \left| \chi \right\rangle.
\label{eq:prooftheorem26}
\end{equation}
Insertion of the obligatory $\hat{R}^2 = \mathds{1}$ factors concludes the proof. $\blacksquare$
\end{appendix}

\bibliographystyle{h-physrev}

\bibliography{bibliography}

\newpage


\end{document}